\documentstyle[preprint,revtex]{aps}
\begin{document}
\draft
\begin{title}
 Interference of two electrons entering a superconductor
\end{title}
\author{F. W. J. Hekking and Yu.V. Nazarov}
\begin{instit}
 Institut f\"ur Theoretische Festk\"orperphysik, Universit\"at Karlsruhe,
 Postfach 6980, 7500 Karlsruhe, FRG\\
\end{instit}

\begin{abstract}
The subgap conductivity of a normal-superconductor (NS) tunnel junction
is thought to be due to tunneling of two electrons.
There is a strong interference between these two electrons,
originating from the spatial phase coherence in the normal metal
at a mesoscopic length scale and the intrinsic coherence of the superconductor.
We evaluated the interference effect on the transport through an NS junction.
We propose the layouts to observe drastic Aharonov-Bohm and Josephson effects.
\end{abstract}

Quantum phase coherence in solids manifests itself basically in two ways.
First, there is an intrinsic coherence in the superconducting state.
In superconductors,
the phase is indeed a macroscopic variable.
This can be observed in a variety of interference experiments, for example,
with tunnel junctions~\cite{Tinkham}.
Second, even in a normal metal electrons are coherent at a mesoscopic length
scale.
Interference between the electrons in a normal metal gives rise to a set of
phenomena which constitutes the subject of a new branch of condensed matter
physics~ \cite{Imry,Altshuler}.

At the NS interface between a normal metal and a superconductor, these two
sources of coherence
may interplay. Very recent experiments \cite{Petrashov} show how interesting
such an interplay may be. These experiments were performed with NS boundaries
of a high transparency.
The phenomenon of Andreev reflection~\cite{Andreev} seems to be responsible for
the peculiarities observed.

In the present paper we focus on the opposite case of a tunnel NS interface.
It is well-known
that an electron with  energy less than $\Delta$, $\Delta$ being
the superconducting energy gap,
can not tunnel to the superconductor and therefore the transport through the
junction is strongly
suppressed at voltages below the gap~\cite{Tinkham}. On the other hand,
two electrons can enter
the superconductor converting into a Cooper pair since this process costs no
energy. Such two-electron
tunneling~\cite{Wilkins,Hekking} determines the subgap conductivity of the
junction. It has been discussed
\cite{We}
that the rate of this two-electron process is often determined by the
interference of the electron
waves on a space scale given by the coherence length,
either in the normal or the superconducting metal. That motivates us to explore
how agents which
act on the phase will influence the subgap conductivity.
Indeed we find the conditions under which a pronounced Aharonov-Bohm and
Josephson
effect can be observed.

Since we consider the low-voltage subgap conductivity, we assume that $T,eV \ll
\Delta$.
Under these condictions the coherence length $L_T =\sqrt{\hbar D/k_BT}$, where
$D$ is
the diffusion constant, is much larger than the one in the superconductor,
$\sqrt{\hbar D/\Delta}$.
Therefore we concentrate on the interference in the normal metal.

The NS interface is described by the Hamiltonian
$
	\hat{H}
	=
	\hat{H}_N + \hat{H}_S + \hat{H}_T
$, where $\hat{H}_N$ and $\hat{H}_S$ refer to the normal and the
superconducting
electrode, respectively.
The tunnel Hamiltonian $\hat{H}_T$ is given by the usual form
$
	\hat{H}_T
	=
	\sum _{\bf{k},\bf{p},\sigma}
	 t_{\bf{kp}}
	 \hat{a}^{\dagger}_{\bf{k},\sigma}
	 \hat{b}_{\bf{p},\sigma} +
         t_{\bf{kp}}^{\ast}
	 \hat{b}^{\dagger}_{\bf{p},\sigma}
         \hat{a}_{\bf{k},\sigma}
$.
Here, operators $\hat{a}_{\bf{k},\sigma}, \hat{b}_{\bf{p},\sigma}$
correspond to the normal and the superconducting electrode,
$t_{\bf{kp}}$ are the tunnel matrix elements which we take to be
spin-independent; the sum is taken over momenta
$\bf{k},\bf{p}$  and spin $\sigma = \uparrow, \downarrow $.
Second order perturbation theory in $\hat{H}_T$ yields the lowest
order contribution to the amplitude of two-electron tunneling.
The $\hat{a}$ operators appearing in this amplitude
remove two electrons from the normal metal electrode with energy
$\xi _k$ and $\xi _{k'}$.
The amplitude thus consists of a sum over intermediate states in the
superconductor
$$
	A_{\bf{k} \uparrow \bf{k}' \downarrow}
	=
	\sum _{p,p'}
	t^*_{\bf{kp}}t^*_{\bf{k'p'}}
	\left\{
	\langle N|
	\hat{b}^{\dagger}_{\bf{p},\uparrow}
	\frac{1}{\xi _{k'} - \hat{H}_S}
	\hat{b}^{\dagger}_{\bf{p'},\downarrow}
	|N-2 \rangle
	\right.
$$
\begin{equation}
	\left.
	-
	\langle N|
	\hat{b}^{\dagger}_{\bf{p'},\downarrow}
	\frac{1}{\xi _{k} - \hat{H}_S}
	\hat{b}^{\dagger}_{\bf{p},\uparrow}
	|N-2 \rangle
	\right\}
\end{equation}
The matrix elements between $|N\rangle$ and $|N-2\rangle$ connect states
differing by two electrons.
In coordinate representation they can be expressed in terms of the usual
anomalous Green's function
$\hat{F}^{\dagger}_{\uparrow\downarrow}(r'_1,r'_2;\omega)$.
Since we assume that $T,eV \ll \Delta$, we consider only $\xi _k,\xi _{k'} \ll
\Delta$
and find
$$
	A_{\bf{k} \uparrow \bf{k}' \downarrow}
	=
	2\pi \int dr_1dr_2dr'_1dr'_2
	\psi _k(r_1) \psi _{k'} (r_2)
	\times
$$
\begin{equation}
	t^*(r_1,r'_1)t^*(r_2,r'_2)
	\hat{F}^{\dagger}_{\uparrow\downarrow}(r'_1,r'_2;\omega=0)
\label{ampl}
\end{equation}
where $\psi _k(r)$ denotes an eigenfunction of an electron
in the normal metal; primed space arguments refer to the superconductor.
In a disordered material, $\psi _k(r)$ and $
\hat{F}^{\dagger}_{\uparrow\downarrow}(r'_1,r'_2;\omega)$
are in general complicated functions depending on the realisation of the
disorder.
Since we are interested only in the interference occurring in the normal metal,
we perform an average of (\ref{ampl}) over states in the superconductor.
This may be done along the lines of~\cite{Averin}; the product of two
tunnel amplitudes will give the normal state conductance $g(r)$
of the tunnel interface per
unit area such that the total conductance $G_T=\int d^2r g(r)$.
The result reads
\begin{equation}
	A_{\bf{k} \uparrow \bf{k}' \downarrow}
	=
	\frac{\hbar}{e^2}
	\frac{\pi}{\nu _N} \int d^2r
	g(r) \psi _k(r) \psi _{k'} (r)
	e^{i\phi (r)}
\end{equation}
where $\phi (r)$ is the
phase of the superconducting condensate
and $\nu _N$ the density of states in the normal metal.
The rate for two-electron tunneling as a function of the applied bias voltage
$V$
is obtained by applying Fermi's Golden Rule.
To obtain the current we have to sum the tunnel rates in both directions.
As a result we find
$$
	I(V)
	=
	\frac{\pi^2 \hbar}{e^3 \nu _N} \int d^2r_1 d^2r_2
	g(r_1) g(r_2) \exp {i(\phi (r_1) - \phi (r_2))} \times
$$
\begin{equation}
	\int d\omega \{f(\omega /2 -eV) -  f(\omega /2 +eV)\}
	\{  P^C_{\omega}(r_1,r_2) +  P^C_{-\omega}(r_1,r_2)      \}
\label{current}
\end{equation}
Here $f$ is the Fermi distribution for electrons
in the normal metal.
Eq.~(\ref{current}) is the central result of our paper, which clearly shows
the interplay between phase coherence in a superconductor
and a normal metal.
The intrinsic coherence of the superconductor is reflected by the appearance
of the phase difference $\phi (r_1) - \phi (r_2)$.
In the normal metal, the two incoming electrons undergo
many elastic scattering events in the junction region before they tunnel
through
the NS interface, leading to interference on a length scale given
by $L_T$~\cite{We}.
These interference effects have been taken into account
by averaging the rate in the standard way~\cite{Altshuler} over
possible scattering events.
The result~(\ref{current}) therefore contains the sum of two Cooperon
contributions $P^C_{\omega}(r,r')$,
which obey the equation~\cite{Stone}
\begin{equation}
\{-\hbar D(\nabla - i2\pi A(r)/\Phi _0)^2  -i\omega\}P^C_{\omega}(r,r') =
\delta(r-r')
\label{Cooperon}
\end{equation}
where $A$ is the vector potential and $\Phi _0 = hc/2e$ the flux quantum.
 From this equation it is clear that the result
does not only depend on properties of the junction (via $G_T$),
but also on its surroundings over a distance $L_T$, due to the interference
occurring on this length scale.

As a simple example we calculate first the subgap conductance
corresponding to a layout where
a  semi-infinite normal wire of thickness $d \ll L_T$ is connected to a
superconducting electrode
by a tunnel junction.
In this case, the only contribution from the spatial integrations
in~(\ref{current})
originates from the tunnel junction at $r_1=r_2=0$.
The solution of (\ref{Cooperon}) for a wire yields
$P^C_{\omega}(0,0) = 1/d\sqrt{-i\omega\hbar D}$. It leads to
\begin{equation}
	G_{wire}
	=
	8\pi ^{5/2}(1-2\sqrt{2})\zeta(-1/2)R_{cor}G_T^2
	\approx
	53.8R_{cor}G_T^2
\label{wire}
\end{equation}
where $R_{cor} = L_T/(e^2\nu _NDd)$ is the resistance of the wire per
correlation length $L_T$.

Interference effects in a mesoscopic system threaded
by a magnetic flux lead to the Aharonov-Bohm effect: the total resistance
depends periodically on the applied flux~\cite{Washburn}.
A similar effect can be observed with the layout depicted in Fig. 1a,
where a small loop with circumference $L$ is inserted into a wire
at distance $l$ of the junction.
The resistance of the loop is denoted by $R_L$; $R_l$
is the resistance of the piece of wire between loop and junction.
The loop is threaded by a magnetic flux $\Phi$.
The conductance~(\ref{current}) at zero temperature ($R_L,R_l \ll R_{cor}$)
reads
\begin{equation}
	G_{loop}
	=
	4\pi ^2G_T^2
	\left\{
	R_l + \frac{R_L}{\sin ^2 \pi \Phi/\Phi _0}
	\right\}
\label{divloop}
\end{equation}
This result is plotted in Fig. 2 (upper curve).
It shows the usual $h/2e$ periodicity~\cite{Washburn} related to the Cooperon.
Moreover it diverges each time when $\Phi = \Phi _0$.
At finite temperature the result~(\ref{wire}) restricts the maximal conductance
by the value $53.8R_{cor}G_T^2$.
But even at zero temperature, when $R_{cor}$ diverges,
the penetration of magnetic flux in the wires that constitute
the loop.
This penetration leads to a shift in the Cooperon energy
$\omega \to \omega + i\alpha ^2(\hbar D/4L^2)(2\pi \Phi /\Phi _0)^2$
where $\alpha = S_{wire}/S_{loop}$ is the ratio of the
area of the wire and the loop.
As a result we obtain instead of (\ref{divloop}):
\begin{equation}
	G_{loop}
	=
	\frac{8\pi ^2G_T^2 R_L(1+2(l/L)\sin ^2 \pi \Phi/\Phi _0)^2}{(\pi \alpha
\Phi/2\Phi _0)(\cos 2\pi \Phi/\Phi _0 +1)
	+ (2\sin ^2\pi \Phi/\Phi _0 )(1+2 (l/L)\sin ^2 \pi \Phi/\Phi _0 )}
\label{loop}
\end{equation}
Thus the divergencies are
removed as can be seen in Fig. 2 (lower curve), where the result is plotted,
taking $l/L =0.5$,
and $ \alpha = 0.1$.
We note that the flux-dependence presented here will be absent when the
superconductor is in the
normal state.

We now turn to a different geometry (Fig. 1b), where instead of a ring
 a fork is attached to the wire, such that we have two tunnel junctions
to the superconductor at different positions $r_1$ and $r_2$.
The subgap conductance will be determined not only by the flux
threading the closed area between the fork and the superconductor,
but also by the magnetic field distribution in the superconductor.
Let us consider the curves $C_N$ and $C_S$ connecting the junctions
1 and 2 in the normal metal and the superconductor, respectively.
The effect will be governed by the phase
$
	\theta
	=
	\phi (r_1)-\phi(r_2)
	+
	(2e/\hbar c)\int_{C_N} \vec{A}.d\vec{x}
$.
In order to obtain a gauge-invariant expression for $\theta$ we
use the relation $\vec{\nabla} \phi - (2e/\hbar c)\vec{A} = \vec{p}_s/\hbar $,
where $\vec{p}_s$ is the momentum of the superconducting condensate.
In this way we arrive at
$
	\theta
	=
	2\pi \Phi /\Phi _0 + \int_{C_S} (\vec{p}_s/\hbar).d\vec{x}
$,
with $\Phi$ the flux penetrating the closed loop formed by
$C_N + C_S$.
This result does not depend on the choice of $C_S$.
It reflects the dependence of $\theta$ on the penetration of
the magnetic field as well as on the
vortex positions in the superconductor.
The effect can be used to monitor these positions.
Introducing the conductances $G_1,G_2$ of the junctions
we obtain
\begin{equation}
	G_{fork}
	=
	53.8
	R_{cor}
	\left[
	G_1^2 + G_2^2 + 2G_1G_2
	\cos \theta
	\right]
\label{fork}
\end{equation}
when $L_T$ is larger than the size of the fork.
The conductance $G_{fork}$ thus combines the
phase coherence in the normal and the superconducting metal.

If we consider tunneling
to a superconductor of finite size,
the transport will be  influenced by charging effects,
if the total capacitance $C$ of the superconducting island
is small enough,
such that the charging energy $E_c = e^2/2C $ is of the order
of $ \Delta$~\cite{Hekking,Eiles}.
We will restrict ourselves to the case $\Delta > E_c$.
In this case the transport to the superconductor will be
due to two-electron processes.
We will present a simple relation between the rate for
these processes and the current which would flow in the absence
of charging effects.
The transfer of two electrons to the superconducting
island increases its electrostatic energy by an amount
$E_{2el}$.
This energy can be changed with the help of an additional
potential $V_g$ applied to the island;
we assume that $E_{2el} \ll E_c, \Delta$.
The first electron entering the superconductor as a quasiparticle
then increases the electrostatic energy of the island by an amount $E_c$.
The energy of the intermediate states in the superconductor
will therefore be shifted by this amount, leading to
a different amplitude
$
	\tilde{A}_{\bf{k} \uparrow \bf{k}' \downarrow}
	=
	(F(E_c)/F(0))
	A_{\bf{k} \uparrow \bf{k}' \downarrow}
$,
with
\begin{equation}
	F(E)
	=
	\frac{4\Delta}{\sqrt{\Delta ^2 - E^2}}
	\arctan \sqrt{\frac{\Delta+E}{\Delta-E}}
\end{equation}
The rate is then given by
$
	2e\tilde{\Gamma}
	=
	(F(E_c)/F(0))^2
	I(E_{2el}/2e)
$.

Finally we investigate a Josephson-like effect, which occurs in the
geometry depicted in Fig. 1c.
In this layout a normal fork is connected to two different superconductors,
to which a small voltage difference $eV_S$ is applied.
The phase difference between the two superconductors and hence
between the junctions at the extensions of the fork will increase linearly with
time:
$\theta = eV_St/\hbar$.
Substituting this phase difference into Eq.~(\ref{fork}) we obtain
\begin{equation}
	G_{J}
	=
	53.8
	R_{cor}
	\left[
	G_1^2 + G_2^2 + 2G_1G_2
	\cos (eV_St/\hbar)
	\right]
\end{equation}
Thus the conductance oscillates with a frequency $\omega _J = eV_S/\hbar$.
The modulation is of the order of the conductance itself.\\

In conclusion we studied the effect of interference on the subgap conductivity
of an NS tunnel junction.
Transport is determined by the transfer of electrons in pairs from the
normal metal to the superconductor.
At low temperatures, interference between the two electrons occurs in the
normal metal
over a longer length scale than in the superconductor.
Therefore, the subgap conductance is determined not only by properties of
the tunnel interface, but also on the layout on the normal side near the
interface over
a distance $L_T$.
These novel interference effects can be made visible by influencing the
electron phase,
e.g. with the help of the Aharonov-Bohm effect or the Josephson effect.
We discuss these effects for various layouts of practical interest, and
present results for the sugap conductance.

\paragraph{Acknowledgements}
The authors are indebted to D. Est\`eve for a very useful discussion which
initiated
this work.
We furthermore want to thank C. Bruder, M. B\"uttiker, H. Schoeller
and G. Sch\"on for discussions.
The financial support of the Deutsche Forschungsgemeinschaft
through SFB 195 is gratefully acknowledged.

\figure{Three geometries discussed in the text: (a) 1-dimensional
normal wire containing a loop
connected to a superconducting electrode by a single junction;
(b) fork geometry connected to the superconductor by two junctions;
(c) fork geometry connected to two different superconductors.}
\figure{Subgap conductance at zero temperature for the geometry of Fig. 1a,
as a function of flux.
Curves correspond to Eqs.~(\ref{divloop}) (upper curve)
and (\ref{loop}) (lower curve)
in the text.}

\end{document}